\documentclass[runningheads]{llncs}

\def\BibTeX{{\rm B\kern-.05em{\sc i\kern-.025em b}\kern-.08em
    T\kern-.1667em\lower.7ex\hbox{E}\kern-.125emX}}

\newcommand{\orcidSRM}	{\href{https://orcid.org/0000-0001-5656-6108}{\protect\includegraphics[scale=0.045]{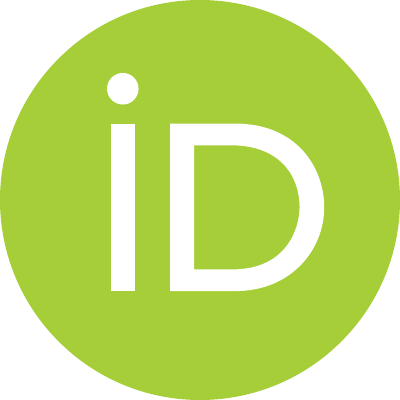}}}

\newcommand{\orcidJB}	{\href{https://orcid.org/0000-0002-3979-400X}{\protect\includegraphics[scale=0.045]{orcid}}}
\usepackage{todonotes}
\usepackage{comment}
\usepackage{makecell}
\usepackage{tabularx}
\usepackage{longtable}
\usepackage{booktabs,cellspace}
\usepackage{wrapfig}
\usepackage{amsmath}
\usepackage{graphicx}
\usepackage{amsfonts}
\usepackage{amssymb}
\usepackage{multirow}
\usepackage{multicol}
\usepackage{algorithm}
\usepackage{algpseudocode}
\usepackage{enumitem}
\usepackage{lipsum}

\usepackage{tablefootnote}
\usepackage{array}
\usepackage[misc]{ifsym}
\usepackage{hyperref}

\newcommand{\pr}{^{\prime}}

\newcommand{\Lt}{\Lambda_{\{\tau\}} }

\newcolumntype{P}[1]{>{\centering\arraybackslash}p{#1}}

\newcommand{\janik}[1]{\textcolor{black}{#1}}
\newcommand{\icpm}[1]{\textcolor{black}{#1}}

\newcommand{\revision}[1]{\textcolor{black}{#1}}
\newcommand{\coopis}[1]{\textcolor{black}{#1}}

\makeatletter
\def\namedlabel#1#2{\begingroup
    #2%
    \def\@currentlabel{#2}
    \phantomsection\label{#1}\endgroup
}
\makeatother

\begin{document}
\title{Collaboration Miner: Discovering Collaboration Petri Nets (Extended Version)}
\titlerunning{Collaboration Miner: Discovering Collaboration Petri Nets}

\author{Janik-Vasily Benzin\inst{1}(\Letter)\orcidJB \and
Stefanie Rinderle-Ma\inst{1}\orcidSRM}
\authorrunning{J.-V. Benzin and S. Rinderle-Ma}
%
\institute{Technical University of Munich, TUM School of Computation, Information and Technology, Garching, Germany\\
\email{\{janik.benzin,stefanie.rinderle-ma\}@tum.de} }

\maketitle              
\begin{abstract}

Most existing process discovery techniques aim to mine models of process orchestrations that represent behavior of cases within one business process. Collaboration process discovery techniques mine models of collaboration processes that represent behavior of collaborating cases within multiple process orchestrations that interact via collaboration concepts such as organizations, agents, and services. While workflow nets are mostly mined for process orchestrations, a standard model for collaboration processes is missing. Hence, in this work, we rely on the newly proposed collaboration Petri nets and show that in combination with the newly proposed Collaboration Miner (CM), the resulting representational bias is lower than for existing models. Moreover, CM can discover heterogeneous collaboration concepts and types such as resource sharing and message exchange, resulting in fitting and precise collaboration Petri nets. The evaluation shows that CM achieves its design goals: no assumptions on concepts and types as well as fitting and precise models, based on 26 artificial and real-world event logs from literature.

\end{abstract}

\keywords{
Collaboration Mining \and Collaboration Process Discovery \and Inter-organizational Processes \and Multi-agent Systems
}
\section{Introduction}
\label{sec:intro}

Process discovery 
aims to discover a 
process model from process executions recorded in an event log \cite{augusto_automated_2019}. 
\icpm{We distinguish two types of executed processes: \emph{Process orchestrations} or \emph{collaboration processes} that represent the control-flow for similar \emph{cases} \cite{diba_extraction_2020} or for \emph{collaborating cases} \cite{benzin_petri_2024} respectively}.
In the first case, \janik{\emph{process orchestration discovery} (POD)} techniques aim at discovering process models of process orchestrations from a set of process instances correlated by cases \cite{diba_extraction_2020}. In the second case, \emph{collaboration process discovery} (CPD) \cite{zeng_top-down_2020,corradini_technique_2022,liu_cross-department_2023} techniques aim to discover a process model of collaboration processes from a set of process instances correlated by collaborating cases. Since collaboration processes are composed of multiple process orchestrations that jointly achieve a shared business goal, its collaborating cases contain multiple cases each corresponding to a particular process orchestration, i.e., cases and orchestrations are in a one-to-one relationship. Most of the CPD techniques target \icpm{their own} class of compositional Petri net to model a collaboration process, \icpm{with exceptions that target \emph{Business Process Modeling Notation}\footnote{\url{https://www.omg.org/spec/BPMN/2.0/}} (BPMN) like \cite{hernandez-resendiz_merging_2021,corradini_technique_2024}, but can be transformed into an equivalent Petri net.}  
Hence, CPD techniques are characterized by targeting different compositional Petri nets, yet a standard model similar to \emph{workflow nets} for process orchestrations is missing \cite{benzin_petri_2024}.

The decision of which model class to target in process discovery is crucial as it determines the \emph{representational bias} \cite{van_der_aalst_representational_2011} that implies the search space for the discovery technique. The variety among CPD techniques results from specializing on certain collaboration processes, e.g., \emph{cross-departmental healthcare processes} (CCHP) in healthcare \cite{liu_cross-department_2023}, \emph{inter-organizational} processes \cite{corradini_technique_2022} in various domains, and \emph{web service compositions} \cite{stroinski_distributed_2019}. These specializations justify assumptions on collaborations and their \emph{interaction patterns} \cite{barros_service_2005}, e.g., only bilateral, point-to-point \emph{message exchanges} exist for \cite{liu_cross-department_2023}. Which of the four collaboration types, i.e., message exchanges, \emph{handover-of-work}, \emph{resource sharing}, and \emph{activity execution}, are supported also originates from the chosen collaboration process, e.g., only message exchanges \revision{and handover-of-work} are discovered in \cite{corradini_technique_2022}. In order to increase generalizability and lower the representational bias of current CPD, we state our research question as follows: \textbf{How can we discover fitting and precise process models of collaboration processes from a \icpm{single event log} in general?} 

By \revision{proposing} \emph{collaboration Petri nets ($cPN$)} and designing the new Collaboration Miner (CM) to discover $cPN$, our contribution results in a generic CPD technique that mines fitting and precise collaboration process models across domains. CM discovers high quality models for all of the 22 artificial event logs that are recorded from multi-agent systems \cite{nesterov_discovering_2023} and inter-organizational processes \cite{corradini_technique_2022}. Moreover, CM discovers high quality models for the four real-world event logs that are recorded from healthcare collaboration processes \cite{liu_cross-department_2023}. As CM with its $cPN$ target supports all four collaboration types and does not assume certain interaction patterns, the representational bias is lowered and model quality is maintained across heterogeneous collaboration processes. \icpm{Note that we assume a single event log recorded from executing a collaboration process is given, i.e., we abstract from event extraction, merging, and correlation \cite{diba_extraction_2020} with corresponding clock synchronization issues as well as privacy concerns \cite{corradini_technique_2024}.} 

\revision{First, basic definitions and notations are repeated in Sect. \ref{sec:fundamentals}. The $cPN$ formalism is introduced in Sect. \ref{sec:cpn}.}
Section \ref{sec:discovery} presents CM by specifying event log requirements 
and a generic approach for discovery. An empirical evaluation of CM in comparison to existing CPD techniques is reported in \autoref{sec:emp}. Next, related work is discussed in \autoref{sec:rel}. Lastly, \autoref{sec:conc} concludes this paper and gives an outlook. 





\section{Preliminaries} 
\label{sec:fundamentals}
We repeat basic definitions and notations.\\
Let X, Y be sets. \\
\noindent$-$ $\mathcal{P}(X)=\left\{X^{\prime} \mid X^{\prime} \subseteq X\right\}$ denotes the \emph{powerset} of $X$, and $\mathcal{P}^+\left(X\right) = \mathcal{P}(X) \setminus \emptyset$ (with $\emptyset$ the empty set) denotes the \emph{set of all non-empty subsets} of $X$. 
Given set $X^{\prime}$, the restriction of $R$'s domain to $X^{\prime}$ is $R_{\mid X^{\prime}}=\{(x, y) \in R \mid x \in X^{\prime}\}$. 
\\
\noindent$-$ A \emph{trace} over $X$ of \emph{length} $n \in \mathbb{N}$ is a function $\sigma:\{1, \ldots, n\} \rightarrow X$. 
For $|\sigma| =0$, we write $\sigma = \epsilon $ and for $|\sigma|>0$, we write $\sigma=\left\langle x_{1}, \ldots, x_{n}\right\rangle$. The \emph{set of all finite traces} over $X$ is denoted by $X^{*}$. 
We write $x \in \sigma$ for $x \in X$, if $\exists_{i \in \{1, \ldots, |\sigma|\}} \; x = \sigma(i) $.
\noindent$-$ A \emph{multiset (or bag)} $m$ over $X$ is a function $m: X \rightarrow \mathbb{N}$, i.e., $m(x) \in \mathbb{N}$ or $x^{m(x)}$ for $x \in X$ denotes the number of times $x$ appears in $m$. For $x \notin X$, we define $ m(x)=0$. $\mathcal{B}(X)$ denotes \emph{the set of all finite multisets} over $X$. 
The \emph{support of multiset} $m \in \mathcal{B}(X)$ is defined by $\operatorname{supp}(m)=\{x \in X \mid m(x)>0\}$, i.e., the support is the set of distinct elements that appear in $m$ at least once. 
We also write $m = [x_1^{m(x_1)}, \ldots, x_n^{m(x_n)}] $ for $\operatorname{supp(m)} = \{x_1, \ldots, x_n \} $. 
Set operations (subset, addition, subtraction) are lifted to multisets in the standard way \cite{van_der_aalst_soundness_2011}. \\
\noindent$-$ Let $\Lambda$ be a finite set of \emph{activity labels}, where $\Lambda_{\{\tau\}} = \Lambda \,\cup\, \tau$ for $\tau \notin \Lambda$ the \emph{silent activity} \cite{barenholz_there_2023}. A \emph{labelled Petri net} is a 5-tuple $ N = (P, T, F, l, \Lt)$, where $P$ is the set of \emph{places}, $T$ is the set of \emph{transitions} with $P \;\cap\; T = \emptyset$, $F \subseteq((P \times T) \cup(T \times P))$ is the \emph{flow} relation, and $l: T \rightarrow \Lt$ is the \emph{transition labelling} function. We define the \emph{preset} of $x \in P \cup T$ by $\bullet x = \{y  \mid (y, x) \in F\}$ and the \emph{postset} of $x$ by $x\bullet =\{y \mid (x, y) \in F\}$. 
A multiset $m \in \mathcal{B}(P)$ is called a \emph{marking}. Given a marking $m$, $m(p)$ specifies the number of tokens in place $p$.
The \emph{transition enabling} $(N, m)[t\rangle$ for $t \in T$ is defined by $(N, m)[t\rangle$ iff $m(p) \geq 1$ for all $p \in \bullet t$. An enabled transition $(N, m)[t\rangle$ can \emph{fire}, \revision{denoted by $(N, m)[l(t)\rangle(N, m\pr)$, resulting in a new marking $m\pr$ defined by $m\pr + \bullet t = m + t \bullet$.} 
\revision{A marking $m\pr$ is \emph{reachable} from $(N, m)$ iff a trace of transition firings exists that starts in $(N, m)$ and ends in $(N, m\pr)$.} 
$N $ is a \emph{workflow net} (WF-net) iff (i) there exists a single source place $i \in P$: 
$\bullet i = \emptyset$; (ii) there exists a single sink place $o \in P$: 
$o \bullet = \emptyset$; and (iii) every node $x \in P \cup T$ is on a directed path from $i$ to $o$. The \emph{initial marking} of $N$ is $[i]$ and the final marking is $[o]$.

\section{Collaboration Petri Nets}
\label{sec:cpn}

This section \janik{introduces} collaboration Petri nets ($cPN$). We conceptualize the organizations/departments, agents, and services that collaborate in a collaboration process by a set of collaboration concepts \revision{$\mathcal{C}$. Each concept's dynamic behavior is a process orchestration. Hence, a \emph{workflow collection} lists the disjunct WF-nets of each collaboration concept in a collaboration process:}

\begin{figure}
  \centering
  \includegraphics[width=0.65\linewidth]{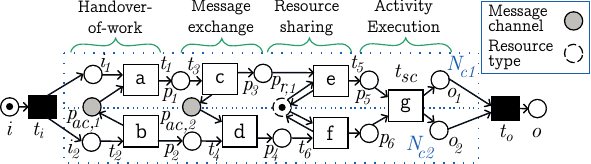}
  \caption{Collaboration Petri net $cPN$ with all four collaboration types.}
  \label{fig:exi} 
  \end{figure}

\begin{definition}[Workflow Collection]
\label{def:gwc}
    \revision{Let $\mathcal{C}$ be the set of collaboration concepts in a collaboration process. A workflow collection is a tuple $WC = (N_c)_{c \in \mathcal{C}}$ of WF-nets $N_{c} = (P_{c}, T_{c}, F_{c}, l_{c}, \Lt)$ with disjunct place and transition names}, i.e., $\forall_{c, c\pr \in \mathcal{C}}$ if $c \neq c\pr$, then $\left(P_{c} \cup T_{c}\right) \cap$ $(P_{c\pr} \cup T_{c\pr}) = \emptyset$. We define:
    \begin{itemize}
        \item $T^{u}=\bigcup_{c \in \mathcal{C}} T_{c}, $ $P^{u}=\bigcup_{c \in \mathcal{C}} P_{c}$, $F^{u}=$ $\bigcup_{c \in \mathcal{C}} F_{c}$, 
        the sets of transitions, places, and arcs of the workflow collection respectively,
        \item $l^u : T^u \rightarrow \Lt$, $l^u(t) = l_c(t) $ for $t \in T_c$.
    \end{itemize}
    
\end{definition}

\icpm{In \autoref{fig:exi}, a $cPN$ is depicted. Two agents ``c1'' and ``c2'' collaborate in this section's running example. \revision{$\mathcal{C}$ equals the two agent names.} Each agent's process orchestration is modelled as a WF-net (denoted in blue in \autoref{fig:exi}) without collaborations.} \coopis{We refer to the process orchestrations in a collaboration process as the ``intra-process'' behavior of the collaboration process.}

There exist four collaboration types $\upsilon \in \Upsilon$: Message exchange ($\upsilon_m$), handover-of-work ($\upsilon_h$), resource sharing ($\upsilon_r$), and activity execution ($\upsilon_s$) \cite{liu_cross-department_2023,benzin_petri_2024}. $\upsilon_m, \upsilon_h, $ and $\upsilon_r$ are asynchronous and $\upsilon_s$ is synchronous. Following existing CPD techniques \cite{liu_cross-department_2023,tour_agent_2023}, handover-of-work is a special case of message exchange. \icpm{For instance, ``c2'' hands the work over to ``c1'' as represented by $p_{ac, 1}$ in \autoref{fig:exi}. 
In contrast, message exchange via asynchronous collaboration places $p_{ac}$ can occur for any transition of a WF-net, e.g., $p_{ac, 2}$.} \icpm{Further collaborations in \autoref{fig:exi} are: Resource sharing of resource type $p_{r,1}$ and an activity execution $t_{sc}$ of activity ``g'' between ``c1'' and ``c2''.} All collaborations between WF-nets of the four types are defined by a \emph{collaboration pattern}:


\begin{definition}[Collaboration Pattern]
\label{def:cp}
    Let $WC = (N_c)_{c \in \mathcal{C}}$ be a workflow collection. A collaboration pattern is a tuple $CP_{WC} = (P_{A C}, P_{RS}, ra, A C, ET)$, where:
    \begin{enumerate}
    \item $P_{A C} $ is the set of asynchronous collaboration places that do not intersect with existing names, i.e., $ P_{A C} \, \cap \, \left(P^{u} \cup T^{u}\right)=\emptyset$ (cf. \autoref{def:gwc}),
    \item $P_{RS} \subseteq P_{A C}$ is the set of shared resource collaboration places,
    \item $ra: P_{RS} \rightarrow \mathbb{N}^+$ is the resource allocation function, i.e., for shared resource type $p_{r} \in P_{RS}$, there exist $ra(p_r)$ shared resources, 
    \item $A C = \{ (p_{ac}, T_s, T_r) \in  P_{A C} \times \mathcal{P}^+(T^{u}) \times \mathcal{P}^+(T^{u}) \mid \forall_{t \in T_s, t^{\prime} \in T_r} \,l^u(t) \neq \tau \wedge l^u(t') \neq \tau \}$ is the asynchronous collaboration relation, i.e., $(p_{ac}, T_{s}, T_{r})$ with $p_{ac} \not\in P_{RS}$ denotes that transitions $t \in T_s$ send a message and transitions $t^{\prime} \in T_r$ receive a message of type $p_{ac}$ via channel $p_{ac}$,
    \item for every $p_r \in P_{RS}$ there exists $ (p_r, T_1, T_2) \in AC $ such that $T_1 = T_2$, i.e., resource types are used and released in transitions $t \in T_1$, and
    \item ET = $\{ (t_{sc}, \,T_{sc}) \in T^u \times \mathcal{P}^+(T^{u}) \mid t_{sc} \in T_{sc} \wedge l^u(t_{sc}) \neq \tau \wedge \forall_{t, t^{\prime} \in T_{sc}} \; l^u(t) = l^u(t')  \}$ is the relation of synchronous collaborations induced by equally-labelled transitions.
    \end{enumerate}
\end{definition}

Observe that all three asynchronuous collaboration types are encoded by relation $AC$ (4). Distinctions are made for resource sharing through $P_{RS}$ (2), the resource allocation $ra$ (3), and the self-loop requirement in (5). Handover-of-work is not explicitly differentiated, since its only difference is the ``location'' of its receiving transitions $T_r$ within the receiving WF-nets. \coopis{Formally, handover-of-work is determined by the condition: given some collaboration concept $c \in \mathcal{C}$, a transition $t \in i_c \bullet$ in the postset of source place $i_c$ in WF-net $N_c$ has an asynchronous collaboration place $p_{ac} \in P_{A C}$ in its preset $p_{ac} \in \bullet t$ \cite{zeng_modeling_2015}.} 
 Inducing synchronous collaboration by equally-labelled transitions (6) follows all existing CPD techniques \cite{nooijen_automatic_2013,zeng_cross-organizational_2013,popova_artifact_2015,van_der_aalst_discovering_2020,nesterov_discovering_2023,liu_cross-department_2023,barenholz_there_2023} with synchronous collaboration.


\coopis{Note that our collaboration pattern builds on existing techniques with respect to collaboration type modeling. The main difference is that we take a global view and we generalize the collaboration types and the message communication models of existing techniques (cf. \autoref{sec:rel}). First, our definition provides a \emph{global view} on the collaborations in a collaboration process, as all collaborations of the collaboration process are defined in a single collaboration pattern. Because the collaboration pattern is separated from the intra-process behavior of the collaboration process, our global view avoids redundancies and simplifies the discovery of collaboration processes in \autoref{sec:discovery}. In contrast, the \emph{local view} of existing techniques (e.g., \cite{liu_cross-department_2023}) represents the collaborations in each concept's process orchestration, i.e., the ``inter-process'' collaboration behavior is included in the ``intra-process'' behavior. Hence, a single collaboration is included multiple times in different process orchestrations such that the collaboration has to be discovered multiple times. Second, our definition generalizes the point-to-point communication model of \cite{corradini_technique_2024} to multiple sender and receivers per message type. Similar to \cite{nesterov_discovering_2023}, our model allows for different sending transitions and receiving transitions per collaboration concept and message type $p_{ac}$. 
}

Given a workflow collection $WC $ and collaboration pattern $CP_{WC}$, a collaboration Petri net is the result of merging the WF-nets in $WC$ as specified by the collaboration pattern in a similar, yet generalized manner to \cite{657557}.

\begin{definition}[Collaboration Petri Net ($cPN$)]
\label{def:cpn}
Let $CP_{WC} = (P_{A C}, P_{RS}, \\ra, A C, ET)$ be a collaboration pattern with $WC = (N_c)_{c \in \mathcal{C}}$ a workflow collection. A Collaboration Petri Net is a marked Petri net $cPN= \biguplus_{c \in \mathcal{C}}^{CP} N_c = ((P, T, F, l, \Lambda_{\{\tau\}}), m_0)$ defined as:
\begin{enumerate}
\item $P=P^{u} \cup P_{A C} \cup\{i, o\} $ (cf. \autoref{def:gwc}),
\item $T=r\left(T^{u}\right) \cup \left\{t_{i}, t_{o}\right\}$, with $r$ a renaming function: $r(x)= t_{sc}$ if there exists a $\left(t_{sc}, T_{sc}\right) \in ET $ such that $t \in T_{sc}$, otherwise $r(x)=x$,
\item $\left\{i, o, t_{i}, t_{o}\right\} \cap\left(P^{u} \cup T^{u} \cup P_{A C} \right)=\emptyset$,

\item $F^{\prime}=F^{u} \;\cup$ \\ $\left\{(t, p) \in T^{u} \times P_{A C} \mid(p, x, y) \in A C \wedge t \in x\right\} \,\cup$ \\$\{(p, t) \in P_{A C} \times T^{u} \mid $$(p, x, y) \in A C \,\wedge \; t \in y\} \;\cup$ $\left\{\left(i, t_{i}\right),\left(t_{o}, o\right)\right\} \;\cup $$\;\{\left(t_{i}, i_{c}\right) \mid c \in \mathcal{C} \} \:\cup\: \{\left(o_{c}, t_{o}\right) \mid c \in \mathcal{C} \}$,

\item $F=\left\{(r(x), r(y)) \mid(x, y) \in F^{\prime}\right\}$,

\item $l(t)= l^u(t)$ if $t \in T^u$, $l(t) = \tau$ otherwise,
\item $m_0(p) = 1$ if $p = i$, $m_0(p) = ra(p)$ if $p \in P_{RS}$ and $m_0(p) = 0$ otherwise.

\end{enumerate} 
\end{definition}

Observe that the example in \autoref{fig:ex} is a $cPN = \biguplus_{c \in \mathcal{C}}^{CP} N_c$. \icpm{The collaboration pattern $CP$ ``consists of'' the two asynchronous message places $p_{ac,1}, p_{ac, 2}$, the asynchronous resource place $p_{r,1}$, and the synchronous activity execution transition $t_{sc}$. The collaboration process starts by instantiating a collaborating case as modelled by $m_0(i) = 1$. The collaborating case corresponds to a case for ``c1'' and to a case for ``c2''. The collaborating case ends after all agent's process orchestrations ended, i.e., the following final marking is reached $m(o) = 1$, $m(p) = ra(p)$ if $p \in P_{RS}$, and $m(p) = 0$ otherwise.}

Note that resource places do not change the semantics in an untimed setting, but $cPN$s extended with time delays for transitions firings would be sensitive to resource places, i.e., discovering resource places supports subsequent analysis. 
\coopis{Also note that the collaboration concepts in a collaboration process interact one-to-one, i.e., there exists a single instance of each collaboration concept for execution. Hence, collaboration processes only intersect with \emph{artifact-} or \emph{object-centric} processes \cite{van_der_aalst_discovering_2020} with respect to synchronous collaboration in a one-to-one relationship. Hence, CPD techniques cannot be generally applied in an object-centric setting.} 
Next, we show how CM discovers $cPN$s from event logs.

\section{Collaboration Miner}
\label{sec:discovery}

CM is a technique to discover a $cPN$ from event log $L$. 
The next section introduces requirements on event logs $L$ such that CM can be applied. In \autoref{sec:alg}, CM with log projection $\pi$ and collaboration discovery $cdisc$ is defined in detail. 

\subsection{Event Log Requirements}
\label{sec:log}

CM takes an event log as input. Event logs are either generated by some collaboration process model, e.g., a $cPN$, or are extracted from information systems that support the process execution \cite{diba_extraction_2020}. We apply the same conceptualization of \emph{interleaving} semantics and totally-ordered traces of events to model business processes as the majority of discovery techniques\janik{, i.e., POD \cite{augusto_automated_2019} and CPD \cite{benzin_petri_2024}}.

\begin{table*}[ht]
\centering
\caption{Five events (represented by rows) of real-world event log $L_{\text{\texttt{EM}}}$ \cite{liu_cross-department_2023}.}
\label{tab:events}
\resizebox{\linewidth}{!}{
\begin{tabular}{lccccccc} 
    \toprule
    Event & case & act (activity)    & timestamp           & c (concept)    &  rs (resource)                         & \makecell{s (send\_msg)} & \makecell{r (receive\_msg)}   \\ 
    \midrule
    $e_1$ & t1 & register & 2019-12-28T00:20:21 &  $\{$Emergency$\}$  &   $\emptyset$         &  $\emptyset$         &  $\emptyset$     \\
    $e_2$ &t1 & rescue & 2019-12-28T01:20:21 &     $\{$Emergency$\}$        &    $\{$charging system$\}$       & $\emptyset$      &    \\
    $e_3$ &t1 & reserve & 2019-12-28T10:20:21 &      $\{$X\_ray$\}$        &    $\{$charging system$\}$         &      $\{$acceptance notice$\}$      &   $\{$reservation form$\}$   \\
    $e_4$ &t1 & plan imaging & 2019-12-28T11:20:21 &      $\{$Surgical$\}$       &    $\emptyset$          &  $\{$photo form$\}$        &  $\{$acceptance notice$\}$     \\
    $e_5$ &t1 & consult & 2019-12-28T23:20:21 &     $\{$Surgical, Cardiovascular$\}$       &   $\{$diagnosis room$\}$           &  $\emptyset$         & $\emptyset$    \\
    \bottomrule
    \end{tabular}}
\end{table*}

\begin{definition}[Event Log]
    \label{def:log}
    Let $\mathcal{A}$ and $\mathcal{V}$ be universes of attribute names and values respectively. An event is a function $e: \mathcal{A} \rightarrow \mathcal{V} $. We denote the universe of events with $\mathcal{E}$. An event log is a multiset of event traces $L \subseteq \mathcal{B}(\mathcal{E}^*)$.
\end{definition}

Table \ref{tab:events} depicts five events $e_1, \ldots, e_5$ as rows with mappings $e_1(\text{case}) = $ t1, $e_2(\text{act}) = $ rescue, $e_3(\text{c} )= \{$X\_ray$\}$,  $e_5(\text{rs}) = \{$diag\-nosis room$\}$ , $e_4(\text{s}) =  \{$photo form$\}$, and $e_4(\text{r}) =  \{$acceptance notice$\}$. All five events $e_1, \ldots, e_5 \in \sigma$ are in the same trace $\sigma \in L_{\text{\texttt{EM}}}$. \icpm{$L_{\text{\texttt{EM}}}$ is recorded from executing a healthcare collaboration process in which hospital departments collaborate to treat patients \cite{liu_cross-department_2023}.}

We distinguish two requirements on event logs. \icpm{Note that we assume a single event log of the collaboration process to be extracted, merged, and correlated already (cf. \autoref{sec:intro}).} 

\begin{enumerate}
    \item[\namedlabel{r1}{\textsl{R1}}] $\forall_{\sigma \in L} \, \forall_{e \in \sigma}\; e(\text{act}) \neq \bot $, i.e., all events of $L$ have a defined activity.
    \item[\namedlabel{r2}{\textsl{R2}}] \janik{$\forall_{\sigma \in L} \, \forall_{e \in \sigma}\; e(\text{c}) \subseteq \mathcal{C} \,\wedge\, (\exists_{e\pr \in \sigma} | e\pr(\text{c}) | \geq 1 \,\lor\, (\exists_{\sigma_1, \sigma_2 \in L, \:e_1 \in \sigma_1, \:e_2 \in \sigma_2} \; e_1(\text{rs}) \,\cap\, e_2(\text{rs}) \neq \emptyset \,\lor\, e_1(\text{s}) \,\cap\, e_2(\text{r}) \neq \emptyset)) $, i.e., all events in the event log $L$ record a set of concepts in the ``c'' (concept) attribute and each trace records \janik{at least} a synchronous collaboration, two traces share a resource, or share a message type.}
\end{enumerate}

POD techniques $disc$ can be applied on event logs that satisfy requirement \ref{r1}. In contrast, CM can only be applied on event logs that satisfy both requirement \ref{r1} and \ref{r2}. \ref{r2} states that each event contains information on the involved collaboration concepts \janik{(attribute ``c''). Additionally, each trace contains a synchronous collaboration, i.e., multiple concepts in the ``c'' attribute, or has at least one shared resource or message type in common with another trace. For instance, $e_4(\text{s}) =  \{$photo form$\}$ in \autoref{tab:events} means that during execution of activity ``plan imaging'' a message of type ``photo form'' is sent. If neither synchronous collaboration, resource sharing, nor message exchanges are recorded, the event log cannot be qualified as recording process executions from collaboration processes.} 
We assume a first-in-first-out message channel per message type with a one-to-one relation between message types and channels similar to \cite{zeng_cross-organizational_2013,liu_cross-department_2023}. Thus, message instance identifiers are not required. 

\coopis{The ``c'' attribute enables applying CPD techniques in general, as otherwise the information on what concept has executed what activity is missing.} Also, the ``c'' attribute enables to discover synchronous collaboration $\upsilon_s$, the ``rs'' attributes enables discovery of resource sharing $\upsilon_r$, and the ``s'' \& ``r'' attributes enable discovery of message exchange $\upsilon_m$ and handover-of-work $\upsilon_h$ collaboration (cf. \autoref{sec:cpn}). Note that an event log recorded from a collaboration process whose collaboration concepts communicate via a \emph{Pub/Sub} \cite{corradini_technique_2024} communication model only meets requirements \ref{r1} and \ref{r2}, if the concepts communicate via messages or another collaboration type, too (cf. \autoref{sec:conc}). 
Also, an event log $L$ that satisfies both requirements can either be serialized into the \emph{eXtensible Event Stream} (XES) log format \cite{noauthor_ieee_2016} or into the \emph{Object-Centric Event Log} (OCEL) format \cite{van_der_aalst_discovering_2020}. Given $L$, \icpm{we can apply CM as proposed in the next section.} 

\subsection{CM Algorithm}
\label{sec:alg}

We start with introducing the log projection $\pi$ \janik{to project event log $L$ on a collaboration concept $c \in \mathcal{C}$}:

\begin{definition}[Log Projection]
    Let $L \subseteq \mathcal{B}(\mathcal{E}^*)$ be an event log that satisfies requirements \ref{r1} and \ref{r2}. Log projection on collaboration concept $c \in \mathcal{C}$ is defined by $\pi_c(L) = [\sigma_{1, c}^{\icpm{L(\sigma_{1})}}, \ldots, \sigma_{n, c}^{\icpm{L(\sigma_{n})}}] $ for $\operatorname{supp}(L) = \{\sigma_1, \ldots, \sigma_n\}$ and $\sigma_{i, c} = \sigma_{i\: |\{e \,\in\, \mathcal{A}\, \rightarrow\, \mathcal{V}\, \mid\, c\, \in\, e(\text{\emph{c}}) \}} $\footnote{\icpm{We inductively define the \emph{projection of a trace} on a set $Y$ by $\epsilon_{\mid Y}=\epsilon,(\langle x\rangle \cdot \sigma)_{\mid Y}=\langle x\rangle \cdot \sigma_{\mid Y}$ if $x \in Y$ and $(\langle x\rangle \cdot \sigma)_{\mid Y}=\sigma_{\mid Y}$ otherwise.}} with $i \in \{1, \ldots, n\}$.
\end{definition}

For example, $\pi_{\text{``Emergency''}}(L_{\text{\texttt{EM}}})$ results in trace $\sigma_1$ to only contain the first two events of the five events depicted in \autoref{tab:events}. 
In the following, we define the Collaboration Miner (CM) \icpm{and illustrate with example event log $L_{\text{\texttt{EM}}}$.}

\begin{figure*}[ht!]
    \centering
    \includegraphics[width=\linewidth]{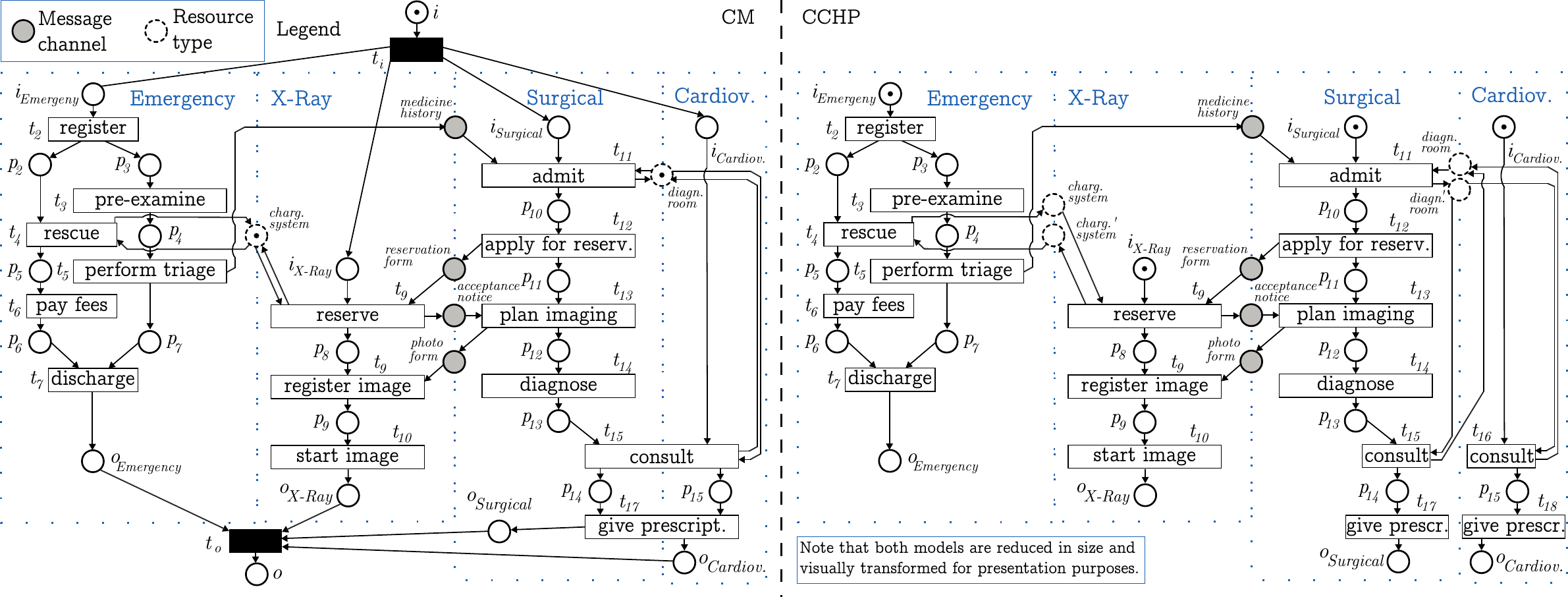}
    \caption{$cPN$ discovered by CM and the composed RM\_WF\_net discovered by CCHP on log \texttt{EM}.}
    \label{fig:ex}
  \end{figure*}

\noindent\namedlabel{step1}{\textbf{Step 1}}. \coopis{Given event log $L$, the first step determines five sets and three functions by extracting attribute information from each event: The set of collaboration concepts $\mathcal{C} = \bigcup_{\sigma \,\in\, L,\; e \,\in \,\sigma} \;e(\text{c}),$ the set of asynchronous message places $ \;P_M = \bigcup_{\sigma \,\in\, L,\; e \,\in \,\sigma} \;e(\text{s})\: \cup\: e(\text{r}), $ the set of asynchronous resource sharing places $ P_{RS} = \bigcup_{\sigma \,\in\, L,\; e \,\in \,\sigma} \;e(\text{rs}), $ and the set of activities $ \:\Lambda_L = \bigcup_{\sigma \,\in\, L,\; e \,\in \,\sigma} e(\text{act.})$. The function $\Lambda_s(x)$ returns the set of activities that sent message $x \in P_M$, $ \Lambda_r(x)$ returns the set of activities that received message $x \in P_M$, and $\Lambda_{rs}(x)$ returns the set of activities that shared resource $x \in  P_{RS}$. All three functions are determined by $ \Lambda_y(x) =  \bigcup_{\sigma \,\in\, L,\; e \,\in \,\sigma, e(y) = x} e(\text{act.})$ for $y \in \{$s, r, rs$\}$ with $x \in P_M$ if $y \neq rs$ and $x \in P_{RS}$ otherwise.} 
\\
\noindent\textbf{Example:} \coopis{For \autoref{tab:events}, we have collaboration concepts $\mathcal{C} = $$\{$Emergency, X-Ray, Surgical, Cardiov.$\}$, asynchronous message places $P_M = $ $\{$photo form, res. form, accept. notice, $\ldots\}$, asynchronous resource sharing places $P_{RS} =$ $\{$charg. system, diagn. room $\}$, the set of activities $\Lambda_L = $ $\{$ register, $\ldots\}$, the function returning sending activities per message $\Lambda_{\text{s}}(x) =$ $\{$(photo form,  $\{$plan imaging$\}$), $\ldots\}$, the function returning receiving activities per message $\Lambda_{\text{r}}(x) =$ $\{$(res. form, $\{$reserve$\}$), (accept. notice, $\{$plan imaging$\}$), $\ldots\}$, and the function returning resource sharing activities per resource $\Lambda_{\text{rs}}(x) =$ $\{$(charg. system,  $\{$rescue, reserve$\}$), $\ldots\}$.}
\\
\noindent\namedlabel{step2}{\textbf{Step 2}}. Project $L$ on collection of event logs $L_{c_1}, \ldots, L_{c_n}$ with $\pi_{c_i}(L) = L_{c_i}$ for $i \in \{1, \ldots, |\mathcal{C}|\}$. Apply POD technique $disc$ on each projected event log $L_{c_i}$ resulting in a collection of WF-nets $N_{c_1}, \ldots, N_{c_n}$. Any POD technique $disc$ can be applied, as long as it discovers WF-nets. Check if a valid WF-net is discovered on each projected event log. Note that if $disc$ discovers \emph{duplicate labels} \cite{augusto_automated_2019}, the respective transitions will be fused as if they represent synchronous collaboration $\upsilon_s$ without an additional label renaming. Construct a workflow collection $WC = (N_{c})_{c \in C}$ with $N_c = (P_c, T_c, F_c, l_c, \Lambda_{L, \{\tau\}}) $ by renaming place and transition names to avoid name clashes.\\
\icpm{\noindent\textbf{Example:} For \autoref{tab:events}, we have $L_{c_1} = $ $\{ e_1, e_2, \ldots\}$, $L_{c_2} = $ $\{ e_3, \ldots\}$, $L_{c_3} = $ $\{ e_4, e_5, \ldots\}$, and $L_{c_4} = $ $\{ e_5, \ldots\}$. We apply \emph{Inductive Miner} \cite{leemans_discovering_2013} as $disc$, resulting in four valid WF-nets $N_{c_1}, \ldots, N_{c_4}$ as highlighted with blue-dotted rectangles on the left in \autoref{fig:ex} (overlapping transitions $t_{15}, t_{17}$ are to be split). Note that the place and transition names are already renamed such that $WC_{ex} = (N_{c})_{c \in \mathcal{C}}$ is a workflow collection.}

\noindent\namedlabel{step3}{\textbf{Step 3}}. Apply collaboration discovery $cdisc$ to mine collaboration pattern $CP$ as defined in the following. Compute sending transitions $T_{\text{s}}(x) = \{t \in T^u \mid l^u(t) \in \Lambda_{\text{s}}(x) \}$ (cf. \autoref{def:gwc}), receiving transitions $T_{\text{r}}(x) = \{t \in T^u \mid l^u(t) \in \Lambda_{\text{r}}(x)\}$, resource sharing transitions $T_{\text{rs}}(x) = \{t \in T^u \mid l^u(t) \in \Lambda_{\text{rs}}(x) \}$, message exchanges $AC\pr = \{ (p_{ac}, T_{\text{s}}(p_{ac}), T_{\text{r}}(p_{ac})) \mid p_{ac} \in P_M \,\wedge\, T_{\text{s}}(p_{ac}) \neq \emptyset \,\wedge\, T_{\text{r}}(p_{ac}) \neq \emptyset \}$, and resource sharing $ AC^{\prime\prime} = \{ (p_{ac}, T_{\text{rs}}(p_{ac}), T_{\text{rs}}(p_{ac})) \mid p_{ac} \in P_{RS}  \,\wedge\, T_{\text{rs}}(p_{ac}) \neq \emptyset \}$. If events in $L$ do not contain information on \emph{lifecycles} \cite{augusto_automated_2019}, set $ra(p_r) = 1 $ for $ p_r \in P_{RS}$, else determine $\operatorname{max}_{p_r}(L)$ the maximum of concurrently running activities sharing $p_r$ and set $ra(p_r) = \operatorname{max}_{p_r}(L)$. Then, collaboration pattern $CP = (P_{RS} \:\cup\: P_M, P_{RS}, ra, AC\pr \,\cup\, AC^{\prime\prime}, ET)$, where $ET$ is induced by equally-labelled transition subsets of all transitions in $WC$ (cf. \autoref{def:cp}). \\
\icpm{\noindent\textbf{Example:} For $WC_{ex}$ (cf. \autoref{fig:ex}), we have sending transitions $T_{\text{s}}(x) = $ \\$\{ (\text{accept. not.}, \{t_9\}), \ldots \}$, receiving transitions $T_{\text{r}}(x) = \{ (\text{res. form}, \{t_{9}\}), \ldots \}$, resource sharing transitions $T_{\text{rs}}(x) = \{ (\text{charg. system}, \{t_4, t_9\}), \ldots \}$, message exchanges $AC\pr = \{ (\text{res. form}, \{t_{12}\}, \{t_{9}\}), \ldots \}$, and resource sharing $ AC^{\prime\prime} = \{(\text{charg. system}, \{t_4, t_{9}\}, \{t_4, t_{9}\}), \ldots \}$. $ra$ is a constant function at value 1, because $L_{\text{\texttt{EM}}}$ does not record lifecycles. Then, $CP = (P_{RS} \;\cup\; P_M, P_{RS}, ra, AC\pr \;\cup\; AC^{\prime\prime}, \{(t_{15}, \{t_{15}, t_{16}\}) \ldots \})$.} \\
\noindent\namedlabel{step4}{\textbf{Step 4}}. Return $cPN = ((P, T, F, l, \Lambda_{L, \{\tau\}}), m_0) = \biguplus_{c \in \mathcal{C}}^{CP} N_c$.\\
\icpm{\noindent\textbf{Example:} The $cPN$ is depicted on the left in \autoref{fig:ex}.}


The parameters of CM are inherited from the respective POD technique $disc$, i.e., no new parameters are added to the algorithm. Similar to existing CPD techniques, CM applies a divide-and-conquer approach on the collaboration concept in the event log and POD technique $disc$ on projected event logs (cf. \ref{step2}), since a collaboration process is a composition of WF-nets. 
Conceptually, CM comes with a general formulation of collaboration concepts
, domain-agnostic event log requirements, no assumptions on the interaction patterns for message exchanges, and supports all four collaboration types (cf. $CP$ in \ref{step3} and \autoref{sec:emp} for details). Consequently, resulting $cPN$s are not specialized on certain collaboration processes. 

Since CM builds on a Petri net theory with activity labels, \icpm{CM supports all $disc$ that discover \emph{duplicate} ($l(t) = l(t\pr)$) and \emph{silent activities} ($l(t) = \tau$). In particular, silent activities are crucial for many control-flow patterns. Both activities are not fused in \ref{step4} (cf. \ref{step2} and \autoref{def:cp}).} Note that $\pi$ projects to the empty trace $\epsilon$ for some $c \in \mathcal{C}$ in \ref{step2}, if a trace does not contain any event with activities executed by $c$. 
The design ensures a fitting $cPN$, as without the ability to ``skip'' a WF-net corresponding traces cannot be perfectly replayed. Thus, this projection conforms to the design goals of CM.

Also, CM still returns a valid $cPN$ in case some parts of the event logs' requirement \ref{r2} are not satisfied: Missing ``rs'' attribute results in $P_{RS}$ to be empty, missing ``s'' with a ``r'' for some message type or vice versa results in the type not to be included (cf. non-empty sets in \ref{step3}), and no ``s'' and ``r'' attributes results in $P_M$ to be empty. However, if the ``c'' attribute is missing, $\mathcal{C}$ is undefined and \ref{step2} cannot be applied, i.e., collaboration concepts must be recorded for every event such that CM discovers a valid $cPN$. 

\subsection{Tool Support}
\label{sec:tool}

The CM implementation is publicly available at \url{https://gitlab.com/janikbenzin/cm} in Python and builds on the PM4PY\footnote{\url{https://github.com/pm4py/pm4py-core}} library, i.e., all POD techniques that discover WF-nets in PM4PY can be applied for $disc$ in CM. 
For the empirical evaluation in the next section, the implementation comes with an automated evaluation pipeline that supports event logs conforming to the XES extensions as defined in \cite{nesterov_discovering_2023} for multi-agent systems, in \cite{corradini_technique_2022} for BPMN collaborations, and in \cite{liu_cross-department_2023} for healthcare collaboration processes. Each of these XES event logs can be automatically converted to the respective event log format required by one of the CPD techniques in \autoref{fig:results}, e.g., XES logs are also converted to equivalent OCEL logs. CM is implemented on the XES extension as defined in \cite{corradini_technique_2022}. From the 14 CPD techniques in \autoref{fig:techniques}, three publicly available CPD technique implementations are packaged with our CM implementation to facilitate reproducing the empirical evaluation automatically. While \emph{object-centric process discovery} (OCPD) \cite{van_der_aalst_discovering_2020} is implemented in PM4PY, the CCHP \cite{liu_cross-department_2023} and \emph{Colliery} \cite{corradini_technique_2022} CPD techniques were originally available with a graphical user interface only. Hence, we have automated both the implementation of CCHP and Colliery such that both CPD techniques are callable via the command line. 
 
\section{Experimental Evaluation}
\label{sec:emp}

\icpm{The 15} CPD techniques depicted in \autoref{fig:techniques} discover different Petri net classes or BPMN diagrams to model collaboration processes from heterogeneous domains with different collaboration types. 
\icpm{From the 15 CPD techniques, only CCHP \cite{liu_cross-organization_2020,liu_cross-department_2023}, Colliery \cite{corradini_technique_2022,corradini_technique_2024}, OCPD \cite{van_der_aalst_discovering_2020}, and Agent Miner \cite{tour_agent_2023} have publicly available implementations and can be applied in our evaluation.} Nevertheless, Agent Miner is excluded from the evaluation for three reasons. 
\revision{First, the supported collaboration type $\upsilon_h$ \revision{together with the assumption that only a single concept can execute an activity simultaneously (cf. Def. 4.1 in \cite{tour_agent_2023}) means that the Agent Miner cannot be applied to event logs with synchronous collaboration. Second, Agent Miner assumes that every directly-follows pair of events $(e_1, e_2)$ with different concept attributes implies a handover-of-work message between the respective concepts in a trace (cf. Def. 4.2 in \cite{tour_agent_2023}}), which is violated in every of the 26 event logs. Third, we still tried to apply Agent Miner on the logs, but were unable to get an output.}  

\begin{table}
\centering
\caption{Overview of existing CPD techniques.}
\label{fig:techniques}
\resizebox{0.7\linewidth}{!}{
\begin{tabular}{lccccc}
\toprule
CPD & Year & Model & $\Upsilon$ & Domains   \\
\midrule
\cite{gaaloul_log-based_2008}  &    2008                       &    WF-nets & $\upsilon_m$ &    Web service    \\
\cite{zeng_cross-organizational_2013} & 2013  & Integrated RM\_WF\_nets & $\upsilon_m, \upsilon_s, \upsilon_r$ & Logistics, Healthcare  \\
\cite{nooijen_automatic_2013,popova_artifact_2015}   & 2013/15           &    Artifact-centric models          &  $\upsilon_s$  &  Accounting     \\
 \cite{stroinski_distributed_2019} & 2019  & Communication nets & $\upsilon_m$ & Web service  \\
 \cite{van_der_aalst_discovering_2020}   &      2020                         &     Object-centric Petri nets  & $\upsilon_s$ & Commerce     \\
  \cite{zeng_top-down_2020} & 2020  & Top-level process model & $\upsilon_m$ & Logistics  \\
  \cite{hernandez-resendiz_merging_2021} & 2021 & BPMN Choreography &  $\upsilon_m$ & Commerce \\
 \cite{fettke_systems_2022} & 2022  & System net & $\upsilon_s, \upsilon_m, \upsilon_h$ & Retail \\
 
   \cite{kwantes_distributed_2022} & 2022  & Industry net & $\upsilon_m$ & Theoretical \\
 \cite{barenholz_there_2023}   &    2023                        &    Typed Jackson nets & $\upsilon_s$ & Commerce      \\
  \cite{nesterov_discovering_2023} & 2023  & Generalized WF-nets & $\upsilon_s, \upsilon_m$ & Multi-agent systems \\
  \cite{tour_agent_2023} & 2023 & Multi-agent system net & $\upsilon_h$ & Healthcare \& other  \\
  \cite{pena_approach_2024} & 2023 & BPMN collab. diagram & $\upsilon_m, \upsilon_h$  & Commerce \\
   \cite{liu_cross-organization_2020,liu_cross-department_2023} & 2023 & Composed RM\_WF\_nets & $\Upsilon$ & Healthcare  \\
   \cite{corradini_technique_2022,corradini_technique_2024} & 2024  & BPMN collab. diagram & \revision{$\upsilon_m, \upsilon_h$} & Healthcare \& other \\
  
\bottomrule
\end{tabular}}
\end{table}

As generation/extraction is out of scope for this paper, event logs from literature are selected. The selection criterion of meeting requirements \ref{r1} and \ref{r2} yields 26 event logs\footnote{Note that the smart agriculture event log in \cite{corradini_technique_2024} only contains signalling between concepts (Pub/Sub) and does not meet the requirements (cf. \autoref{sec:log}).}: 12 artificial event logs from multi-agent systems \cite{nesterov_compositional_2021,nesterov_discovering_2023}, 10 artificial event logs from BPMN collaborations \cite{corradini_technique_2022}, and 4 real-world event logs from healthcare collaboration processes \cite{liu_cross-department_2023}. \janik{For 17/22 artificial event logs (\texttt{1}-\texttt{A5} in \autoref{fig:results}), the \emph{true} process models that generated the respective event log are available or convertible by PM4PY's ``BPMN to Petri net'' (see True in \autoref{fig:results}). For the five artificial event logs \texttt{R1}-\texttt{R5} conversion with PM4PY is not possible due to the BPMN model structure, so their true process models are not available. Also, true process models are not available for real-world event logs.} 
Descriptive statistics of the 26 event logs are reported in \autoref{fig:results}. The event logs vary along the dimensions: \# of events, average trace length, \# of collaboration concepts, collaboration types $\upsilon$, and properties describing the respective interaction pattern of message exchanges $\upsilon_m$ \cite{barros_service_2005}. Interaction patterns vary along the maximum number of collaboration concepts interacting through a message type, i.e., \emph{one}-way and \emph{two}-way bilateral or \emph{multi}lateral interactions, the maximum number of transmissions per type, i.e., \emph{single} or \emph{multi}, and the relation between activities sending/receiving messages per type, i.e., point-to-point (denoted by 1:1 in \autoref{fig:results}), one-to-many (1:n), and many-to-many interaction (m:n).

Each of the 26 event logs is converted to the respective \janik{CPD technique's} event log input format. 
We apply CM, CCHP, Colliery, and OCPD on each converted event log with the Inductive Miner \cite{leemans_discovering_2013} for POD $disc$ to ensure result differences being caused by the CPD techniques properties, e.g., supported collaboration types, and design choices, e.g., assumptions on the interaction pattern. Inductive Miner is chosen for its formal guarantee to discover perfectly fitting WF-nets such that this design goal can be achieved. \revision{We report the model \emph{size} as the sum of the \# of places and \# of transitions in \autoref{fig:results}.} 
We apply \emph{alignment-based fitness} and \emph{precision} \cite{van_der_aalst_replaying_2012,adriansyah_measuring_2015} to measure model quality with PM4PY except for logs \texttt{ID} to \texttt{SD} in \autoref{fig:results} for which we manually computed fitness and precision (annotated with $^*$) using the more efficient ProM plugin with similar parameters, as PM4PY exceeds a space limit of $>$58GB on a Fedora machine with 14-core Intel i5-13500T (13th Gen) CPU and 64 GB RAM. 

\cite{polyvyanyy_monotone_2019,polyvyanyy_monotone_2020} propose monotone alternatives to both alignment-based metrics. However, neither version of the monotone alternatives is suitable for our evaluation. First, the perfectly-fitting version results in metrics equal to zero as soon as no trace in the event log can be replayed by the discovered model. If a CPD technique does not support synchronous collaboration $\upsilon_s$, e.g., Colliery, it discovers a model with duplicate labels for synchronous collaboration $\upsilon_s$ due to projection. As logs may contain only traces with $\upsilon_s$, metrics are zero and too low. For example, \revision{7/8 logs with $\upsilon_s$ contain only traces with $\upsilon_s$} (\texttt{9}-\texttt{11} \revision{and \texttt{EM}-\texttt{SD}} in \autoref{fig:results}). Second, the partially-fitting version often exceeds a heap space limit of 58GB ($>$3x the space of \cite{augusto_automated_2019}). Third, alignments allow arbitrary final markings. In contrast, the monotone metrics do not take the final marking as input\footnote{\url{https://github.com/jbpt/codebase/tree/master/jbpt-pm/entropia}}, but compute it based on the assumption that places in the final marking are sink places. CCHP requires a final marking for resource places that are neither a sink place nor self-loops \cite{liu_cross-department_2023}, so resource places cannot be removed (cf. \autoref{sec:cpn}) and monotone metrics cannot be computed for CCHP models with resource sharing. Altogether, alignment-based metrics are more suitable, as they can handle partially-fitting traces, are more efficient in terms of heap space, and more flexible in terms of final markings.

\icpm{We illustrate differences of CM and CCHP with their Petri nets discovered on $L_{\text{\texttt{EM}}}$. CCHP is ``closest'' to CM, as CCHP similarly claims to discover all four collaboration types (cf. \autoref{fig:techniques}) and applies a comparable approach to discovery. Hence, the illustration supports understanding the following different results. In \autoref{fig:ex}, the $cPN$ discovered by CM (CM's model) and on the right the composed RM\_WF\_net (CCHP's model) is depicted. While CM's model has a global source and sink place, CCHP does not. Hence, CM's model has the notion that a token in $i$ represents a collaborating case. Most parts of the two models are similar, as CCHP is proposed using $L_{\text{\texttt{EM}}}$, i.e., $L_{\text{\texttt{EM}}}$ does not violate any assumptions by CCHP. CCHP's model has two problems. First, it does not discover $\upsilon_s$ as claimed, since it discovers duplicate activities for ``consult'' and ``give prescription''. Second, it does not discover self-loop places for resources and no resource allocation such that ``Surgical'' and ``Cardiov.'' can never reach their sink place. With marked resource places, CCHP's model would force either ``Surgical'' or ``Cardiov.'' to execute ``consult''. 
}

\begin{table*}[htbp!]
\caption{Model quality metrics of the true model, if available, and discovered by CM, CCHP \cite{liu_cross-department_2023}, Colliery \cite{corradini_technique_2022}, and OCPD \cite{van_der_aalst_discovering_2020} based on artificial event logs 1-12 \cite{nesterov_compositional_2021,nesterov_discovering_2023}, A1-A5 \& R1-R5 \cite{corradini_technique_2022}, and real-world event logs EM-SD \cite{liu_cross-department_2023}, where $\Upsilon_{\sigma} = \{ \upsilon_x \mid x \in \sigma \}. $}
\label{fig:results}
\centering
\resizebox{1\columnwidth}{!}{
\begin{tabular}{@{}ll*{13}{c}@{}}
\toprule
\multicolumn{2}{l}{Event log \( L \)}  & 1 & 2 & 3 & 4 & 5 & 6 & 7 & 8 & 9 & 10 & 11 & 12 & A1 \\
\cmidrule(l{2em}){1-15}
\multicolumn{2}{l}{\hspace{0.1cm}\# Events}   & 95052 & 149988 & 92668 & 102404 & 182452 & 123322 & 88068 & 157098 & 115000 & 102548 & 160000 & 88089 & 100 \\
\multicolumn{2}{l}{\hspace{0.1cm}Avg. trace length}    & 19 & 30 & 19 & 20 & 36 & 25 & 18 & 31 & 23 & 21 & 32 & 18 & 8\\
\multicolumn{2}{l}{\hspace{0.1cm}\# Col. concepts}  & 2 & 2 & 2 & 2 & 2 & 2 & 2 & 3 & 2 & 2 & 2 & 2 & 2 \\
\multicolumn{2}{l}{\hspace{0.1cm}Col. types \(\upsilon\)} & \(\upsilon_m\) & \(\upsilon_m\) & \(\upsilon_m\) & \(\upsilon_m\) & \(\upsilon_m\) & \(\upsilon_m\) & \(\upsilon_m\) & \(\upsilon_m\) & \(\Upsilon_{\langle m, s\rangle}\) & \(\Upsilon_{\langle m, s\rangle}\)  & \(\Upsilon_{\langle m, s\rangle}\)  & \(\Upsilon_{\langle m, s\rangle}\)  & \(\upsilon_m\) \\
\multicolumn{2}{l}{\hspace{0.1cm}\(\upsilon_m \text{: Max. col. con.}\)} & one & one & one & two & two & two & two  & multi & two  & two   & two  & two & one \\
\multicolumn{2}{l}{\hspace{0.1cm}\(\upsilon_m \text{: Max. trans.}\)} & single & single & single & single & single & single & multi  & multi & single  & single   & single &   single  & single \\
\multicolumn{2}{l}{\hspace{0.1cm}\(\upsilon_m \text{: Activity rel.}\)} & 1:n & m:n & 1:1 & m:n & 1:n & 1:n & m:n  & m:n & m:n  & m:n   & 1:n  & 1:n & 1:1 \\
\midrule
\multirow{2}{*}{True} 
 & Precision & 0.716 & 0.401 & 0.754 & 0.759 & 0.39 & 0.564 & 0.817 & 0.481 & 0.714 & 0.793 & 0.495 & 0.766 & 0.972\\
 & Size & 66 & 100 & 88 & 76 & 109 & 113 & 61 & 128 & 105 & 78 & 94 & 86 & 22  \\
\midrule
\multirow{3}{*}{CM} & Fitness & \textbf{1.0} & \textbf{1.0} & \textbf{1.0} & \textbf{1.0} & \textbf{1.0} & \textbf{1.0} & \textbf{1.0} & \textbf{1.0} &  \textbf{1.0} &  \textbf{1.0} & \textbf{1.0} &  \textbf{1.0}  & \textbf{1.0}  \\
& Precision& 0.736 &  0.351 & \textbf{0.765} & \textbf{0.792} &  0.208 &  0.586 & \textbf{0.817} & \textbf{0.504} & \textbf{0.716} &  \textbf{0.781} &  0.433 &  \textbf{0.758} & \textbf{0.972} \\
& Size & 75 & 125 & 103 & 92 & 167 & 132 & 77 & 148 & 120 & 88 & 125 & 95 & 26  \\
\cmidrule(l){2-15}
\multirow{3}{*}{CCHP} & Fitness & ex & ex & \textbf{1.0} & ex & ex & ex & ex & ex & ex & ex & ex & 0.384  & \textbf{1.0} \\
& Precision & ex & ex & \textbf{0.765} & ex & ex & ex & ex & ex & ex & ex & ex & 0.583 & \textbf{0.972}  \\
& Size & 78 & 127 &  \textbf{95} & 92 & 165 & 135 & 78 & 150 & 121 & 94 & 128 & 99 & 26  \\
\cmidrule(l){2-15}
\multirow{3}{*}{Colliery} & Fitness& 0.867 &  0.966 &  \textbf{1.0} &  0.71 &  0.964 &  0.924 &  0.667 &  0.699 &  0.641 &  0.707 &  0.93 &  0.893 & \textbf{1.0} \\
& Precision & \textbf{0.738} & \textbf{0.426} & \textbf{0.765} &  0.747 &  \textbf{0.26} & \textbf{0.598} &  0.686 &  0.306 &  0.5 &  0.67 &  \textbf{0.465} &  0.697 & \textbf{0.972} \\
& Size & 79 & 131 & 103 & 94 & 169 & 136 & 77 & 295 & 130 & 96 & 128 & 104 & 26\\

\cmidrule(l){2-15}
\multirow{3}{*}{OCPD} & Fitness & \textbf{1.0} & \textbf{1.0} & \textbf{1.0} & \textbf{1.0} & \textbf{1.0} & \textbf{1.0} & \textbf{1.0} & \textbf{1.0} & \textbf{1.0} & \textbf{1.0} & \textbf{1.0} & \textbf{1.0}   & \textbf{1.0} \\
& Precision& 0.714 & 0.341 & 0.719 & 0.701 & 0.136 & 0.545 & 0.594 & 0.387 & 0.677 & 0.726 & 0.185 & 0.717 & 0.753 \\
  & Size  &  \textbf{74} &  \textbf{123} &  101 &  \textbf{90} &  \textbf{121} &  \textbf{128} &  \textbf{74} &  \textbf{142} &  \textbf{118} &  \textbf{86} &  \textbf{106} &  \textbf{93} &  \textbf{24}\\
\bottomrule
\end{tabular}
}
\resizebox{1\columnwidth}{!}{
\begin{tabular}{@{}ll*{13}{c}@{}}
\toprule
\multicolumn{2}{l}{Event log \( L \)}  & A2 & A3 & A4 & A5 & R1 & R2 & R3 & R4 & R5 & EM & ID & FP & SD \\
\midrule
\cmidrule(l{2em}){1-15}
\multicolumn{2}{l}{\hspace{0.1cm}\# Events}   & 100 & 100 & 100 & 100 & 22 & 100 & 100 & 100 & 100 & 18909 & 50427 & 37816 & 4320\\
\multicolumn{2}{l}{\hspace{0.1cm}Avg. trace length}   & 18 & 23 & 6 & 24 & 7 & 15 & 18 & 18 & 13 & 32 & 25 & 25 & 23 \\
\multicolumn{2}{l}{\hspace{0.1cm}\# Col. concepts}  & 2 & 3 & 2 & 4 & 2 & 2 & 3 & 3 & 3 & 6 & 6 & 4 & 4  \\
\multicolumn{2}{l}{\hspace{0.1cm}Col. types \(\upsilon\)}  & \(\upsilon_m\) & \revision{\(\Upsilon_{\langle m, h\rangle}\)} & \(\upsilon_m\) &  \revision{\(\Upsilon_{\langle m, h\rangle}\)} & \(\upsilon_m\) & \(\upsilon_m\) & \revision{\(\Upsilon_{\langle m, h\rangle}\)} & \revision{\(\Upsilon_{\langle m, h\rangle}\)} & \revision{\(\Upsilon_{\langle m, h\rangle}\)} &  \(\Upsilon\) & \makecell[c]{\(\Upsilon_{\langle m, s, h\rangle}\) } & \makecell[c]{\(\Upsilon_{\langle m, s, h\rangle}\) } & \makecell[c]{\(\Upsilon_{\langle m, s, h\rangle}\) } \\
\multicolumn{2}{l}{\hspace{0.1cm}\(\upsilon_m \text{: Max. col. con.}\)}& two & one & two & two & two & two & two & two & two & two & two &  two & two \\
\multicolumn{2}{l}{\hspace{0.1cm}\(\upsilon_m \text{: Max. trans.}\)} & single & single & single & single & single & multi & multi & single & single & single   &single  &single  & single  \\
\multicolumn{2}{l}{\hspace{0.1cm}\(\upsilon_m \text{: Activity rel.}\)} & 1:1 & 1:1 & 1:1 & 1:1 & 1:1 & 1:1 & 1:1 & 1:1 & 1:1 & 1:1 & 1:1 & 1:1 & 1:1 \\

\midrule
\multirow{2}{*}{True} 
 & Precision & 0.648 & 0.645 & 0.998 & 0.646 & n/a & n/a & n/a & n/a & n/a & n/a & n/a & n/a & n/a \\
& Size & 47 & 59 & 18 & 63 &  n/a &  n/a &  n/a &  n/a &  n/a &  n/a &  n/a &  n/a &  n/a \\  
\midrule
\multirow{3}{*}{CM} & Fitness & \textbf{1.0} & \textbf{1.0} & \textbf{1.0} & \textbf{1.0} & \textbf{1.0} & \textbf{1.0} & \textbf{1.0} & \textbf{1.0} & \textbf{1.0} & \textbf{1.0} & \textbf{1.0} & \textbf{1.0} & \textbf{1.0} \\
& Precision  &  0.606 &  0.435 & \textbf{0.998} &  0.44 & \textbf{0.538} &  0.419 &  0.293 &  0.649 & \textbf{0.945} & \textbf{0.986} & \textbf{0.797}$^*$  & \textbf{0.789}$^*$ & \textbf{0.845}$^*$ \\
& Size & 50 & 64 & 22 & 75 & 20 & 42 & 80 & 54 & 71 & 95 & 74 & 74 & 65 \\
\cmidrule(l){2-15}
\multirow{3}{*}{CCHP} & Fitness &  0.935 &  0.958 & \textbf{1.0} &  0.905 & \textbf{1.0} &  0.894 &  0.974 &  0.889 &  0.844 & ex & \textbf{1.0} &  0.943 &  0.979 \\ 
& Precision & \textbf{0.648} &  0.467 & \textbf{0.998} &  0.471 & \textbf{0.538} &  0.422 & \textbf{0.369} & \textbf{0.758} & \textbf{0.945} & ex & \textbf{0.797}$^*$  & 0.615$^*$  & 0.777$^*$ \\
& Size & 51 & 65 & 22 & 71 & 20 & 41 & 84 & 55 & 61 & 105 & 74 & 76 & 66 \\
\cmidrule(l){2-15}
\multirow{3}{*}{Colliery} & Fitness &   0.935 &  0.935 & \textbf{1.0} &  0.888 &  0.6 &  0.894 &  0.846 &  0.757 &  0.785 &  0.918 &  0.865 &  0.88 &  0.857 \\
& Precision  & \textbf{0.648} & \textbf{0.645} & \textbf{0.998} & \textbf{0.646} &  0.333 & \textbf{0.439} &  0.333 &  0.638 &  0.814 &  0.253 & 0.233 & 0.176 & 0.229 \\
  & Size & 51 & 63 & 22 & 67 &  \textbf{12} & 43 & 82 & 48 & 61 & 81 &  \textbf{62} & 67 &  \textbf{51} \\
\cmidrule(l){2-15}
\multirow{3}{*}{OCPD} & Fitness & \textbf{1.0} & \textbf{1.0} & \textbf{1.0} &  0.971 & \textbf{1.0} &  \textbf{1.0} &  0.995 &  \textbf{1.0} &  0.868 &  \textbf{1.0} & \textbf{1.0} & \textbf{1.0} & \textbf{1.0} \\
& Precision &  0.598 &  0.399 &  0.703 &  0.314 & \textbf{0.538} &  0.253 &  0.193 &  0.201 &  0.251 &  0.148 & 0.409$^*$ & 0.530$^*$ & 0.509$^*$  \\
  & Size &  \textbf{48} &  \textbf{60} &  \textbf{20} &  \textbf{65} & 20 &  \textbf{30} &  \textbf{66} &  \textbf{41} &  \textbf{59} &  \textbf{76} & 68 &  \textbf{66} & 58  \\
\bottomrule
\end{tabular}
}
\end{table*}

Considering fitness, CM is the only CPD technique that always discovers a $cPN$ that perfectly fits event log $L$ (cf. \autoref{fig:results}). CCHP discovers Petri nets whose final markings are unreachable for event logs \texttt{1}, \texttt{2}, and \texttt{4}-\texttt{11}, as CCHP implicitly assumes a point-to-point (1:1) activity relation, but still adds an interaction for each unique activity pair, e.g., a 2:2 relation yields four activity pairs. As resource places are not discovered as self-loop places by CCHP (cf. \autoref{fig:ex}), the only log \texttt{EM} with resource sharing results in a Petri net with unreachable final marking. Since support for synchronous collaboration $\upsilon_s$ is not implemented in CCHP, Petri nets discovered on event logs with $\upsilon_s$ have a lower fitness similar to Colliery that does not support $\upsilon_s$ by design. \icpm{Because OCPD only supports $\upsilon_s$ (cf. \autoref{fig:techniques}), discovered Petri nets correspond to the parallel execution of all collaboration concept WF-nets for event logs without $\upsilon_s$. If $\upsilon_s$ is contained in an event log, the parallel branches that correspond to a collaboration concept WF-net are synchronized by synchronous collaborations.} Instead of projecting logs, OCPD \emph{flattens} \cite{van_der_aalst_discovering_2020} the log such that the empty trace $\epsilon$ can never be in a flattened event log. Consequently, a concept's WF-net can never be skipped, but skipping is required for logs \texttt{A5}, \texttt{R3}, and \texttt{R5}. Therefore, fitness is not perfect for these logs. \janik{Also, CCHP and Colliery do not allow $\epsilon$ in projected event logs such that CM is the only technique that achieves perfect fitness on logs \texttt{A5}, \texttt{R3}, and \texttt{R5}.}

Considering precision, CM and Colliery discover the most precise models for \janik{15/26 and 12/26} event logs respectively, i.e., either CM or Colliery discover the most precise model with the exception of CCHP for \texttt{R3} and \texttt{R4}. In particular, the most precise model is in the same range as the true model's precision. CM can discover most precise models across all three different event log groups and, thus, across multi-agent systems, inter-organizational processes and healthcare collaboration processes. In particular, CM discovers most precise models for the four real-world event logs \texttt{EM}-\texttt{SD} that contain all or the majority of collaboration types. Colliery does not discover precise Petri nets for the four real-world event logs, because these event logs record many collaborations that are not supported by Colliery. Considering subpar precision metrics of CM, results are often close to the best precision, e.g., event log \janik{\texttt{1} with 0.736 vs. 0.738} or \texttt{5} with 0.586 vs. 0.598
. CCHP discovers the most precise model for 9/26 event logs due to the final marking being unreachable for 11/26 event logs. OCPD usually discovers the least precise model due to the low precision of \janik{parallelly executed} WF-nets \janik{without any message exchange}. 

\coopis{Considering size, OCPD regularly discovers the smallest model, since it does not discover asynchronous places. CCHP, Colliery, and CM are typically in the same range of size. An exception is the event log \texttt{1} for which Colliery discovers a model of twice the size discovered by other CPD techniques. Overall, the sizes of discovered models are close to each other and usually in the range of 1.2x the true model size.}

To sum up, the results show that support for synchronous collaboration increases fitness, support for asynchronous collaboration increases precision, and violated assumptions on interaction patterns significantly decrease fitness and precision and can lead to models in which the final marking is not reachable. 
The experimental evaluation with 26 artificial and real-world event logs with a diverse set of collaboration processes, collaboration types, and interaction patterns shows that CM achieves its design goals of precise and fitting process models without assumptions on concepts, types or patterns.

\section{Related Work}
\label{sec:rel}

\icpm{
To start with, we elaborate on our differences in detail to CCHP \cite{liu_cross-organization_2020,liu_cross-department_2023} that is the CPD technique closest to CM. 
CCHP \cite{liu_cross-organization_2020,liu_cross-department_2023} shows multiple inconsistencies between paper and implementation; hence
we use \namedlabel{cchp}{CCHP sources} \url{https://github.com/promworkbench/ShandongPM/} as substitute
for parts that are undefined, e.g., $cdisc$.}
\icpm{While CCHP has a similar divide-and-conquer approach and modelling of collaboration types, it differs in several aspects in which CM improves CCHP as shown in \autoref{sec:emp}.
CCHP does not allow empty traces in projected logs, which lead to reduced fitness. It assumes a one-to-one activity relation per message channel, resulting in unreachable final markings.
$\upsilon_s$ is only theoretically discovered and $\upsilon_r$ is practically not discovered as self-loop places. 
CCHP does not discover resource allocations and does not specify event log requirements, which leaves an early decision of applicability to the user. Lastly, CCHP is not defined with activity labels such that duplicate and silent activities are always fused. Hence, POD techniques such as the Inductive Miner lead to undesirable results.} 
Overall, CM generalizes, improves, and extends CCHP.

\icpm{
Regarding collaboration types, this work provides a global view on the collaboration process to reduce redundancies and separates the process orchestrations (intra-process) from the collaborations (inter-process) to simplify the discovery of collaboration processes. 
Not separating intra-process from inter-process behavior as done in the RM\_WF\_nets \cite{zeng_cross-organizational_2013} requires to simultaneously discover both behaviors for each collaboration concept. As collaborations are nonetheless discovered, 
collaborations are discovered multiple times. }

\icpm{In the following, we give a quick overview on a selection of the remaining 14 CPD techniques (cf. \autoref{fig:techniques}).} \cite{gaaloul_log-based_2008} discover web service compositions by conceptualizing each message exchange as a single activity. \icpm{\cite{zeng_cross-organizational_2013} discover inter-organizational processes without choices.}  
\cite{stroinski_distributed_2019} propose to discover hierarchically structured services that collaborate via message exchanges. \cite{zeng_top-down_2020} discover \emph{top-level process models} in which the inter-process level is discovered and, subsequently, refined with \emph{local} process models. \icpm{\cite{hernandez-resendiz_merging_2021} discover process choreographies with a focus on issues during merging of distributed event logs.} \cite{corradini_technique_2024,corradini_technique_2022} discover BPMN collaboration diagrams that are the result of converting discovered WF-nets $N_c$ for each partner to BPMN and connecting activities with \emph{message flows}. \coopis{\cite{corradini_technique_2024} is the only CPD technique that supports a \emph{Pub/Sub} communication model in which the collaboration concepts communicate via \emph{signals} and a sent signal can be received multiple times by different concepts.} \cite{kwantes_distributed_2022} formally analyse the extent to which POD techniques $disc$ can be applied to discover models of asynchronously collaborating systems. \cite{nesterov_discovering_2023} propose to discover a \emph{generalized} WF-net (WF-nets that have multiple source and sink places) for each agent that collaborate via message exchanges and synchronous activity execution with each other using a soundness-guaranteeing PD technique $disc$ such as the Inductive Miner \cite{leemans_discovering_2013}. Next, \cite{tour_agent_2023} propose to discover a \emph{multi-agent system net} (MAS) net, which is a WF-net with labels that include the activity label and the agent executing the activity. 
The agents in a MAS net collaborate via handover-of-work. In general, CM advances CPD techniques in the direction of a generalized formulation, fewer assumptions, discovery of resource allocations, support of all four collaboration types, of silent activities, and of more POD techniques $disc$.

Due to CM's goal to discover models of collaboration processes and their one-to-one correspondence between process instances of collaboration concepts, it cannot be equally applied to event logs with one-to-many or many-to-many relationship between process instances of collaboration concepts or more specifically object types. Consequently, CM is only related to CPD techniques \cite{nooijen_automatic_2013,popova_artifact_2015,van_der_aalst_discovering_2020,fettke_systems_2022,barenholz_there_2023} on event logs satisfying requirements \ref{r1}, \ref{r2}, and no multiplicities between collaboration concept identifiers exist. \icpm{These CPD techniques aim to integrate the control-flow and data perspective and have, thus, a different goal.}
Due to the intersection on certain event logs, we still subsume \cite{nooijen_automatic_2013,popova_artifact_2015,van_der_aalst_discovering_2020,fettke_systems_2022,barenholz_there_2023} under CPD techniques in this paper despite their different goal.  
\cite{nooijen_automatic_2013,popova_artifact_2015} discover \emph{artifact-centric} process models consisting of multiple artifact lifecycle models discovered from Enterprise Resource Planning systems. \cite{fettke_systems_2022} delineate a discovery framework with \emph{true concurrent} semantics that does not require a total order on events in the event log. \cite{barenholz_there_2023} propose a framework for (re-)discovering \emph{typed Jackson nets} that are a block-structured, sound-by-design subclass of \emph{typed Petri nets with identifiers}. 

\section{Conclusion and Outlook}
\label{sec:conc}

We propose Collaboration Miner (CM) to discover fitting and precise process models of collaboration processes. Among the heterogeneous set of existing target model classes proposed to model collaboration processes, CM \revision{proposes} collaboration Petri nets ($cPN$). The $cPN$ target class along with the CM's divide-and-conquer approach on the event log and custom collaboration discovery $cdisc$ lowers the representational bias between discovered models and the true collaboration process. More specifically, CM generalizes over domains and their collaboration processes through collaboration concepts, types, and event log requirements. In addition, $cdisc$ eliminates assumptions of existing techniques on the interaction patterns in the event log. CM's ability to discover high-quality models across domains and interaction patterns is empirically shown on 26 artificial and real-world event logs. Future directions are towards providing soundness guarantees, discovering multiplicities between collaboration concept instances, and \coopis{supporting the Pub/Sub communication model for collaboration type $\upsilon_m$. Extending CM to also discover process models that can represent multiplicities between collaboration concept instances necessitates an extension to $cPN$s that would bring collaboration and object-centric process mining closer together.}

\end{document}